\documentclass[pra,notitlepage,showpacs,showkeys]{revtex4-2}
\usepackage{graphicx}
\usepackage{amsmath}
\usepackage{amssymb}

\usepackage[pdfauthor={Richard J. Mathar}, colorlinks=true,
urlcolor=blue,
pdfkeywords={{Foucault Pendulum},{Lagrangian},{Theoretical Mechanics}},bookmarksopen=true,pdfpagelayout=OneColumn]{hyperref}

\begin{document}

\title[Foucault Pendulum]{The Foucault Pendulum: Trajectories of the Full Lagrangian}

\author{Richard J. Mathar}
\homepage{https://www.mpia-hd.mpg.de/homes/mathar}
\affiliation{Max-Planck Institute of Astronomy, K\"onigstuhl 17, 69117 Heidelberg, Germany}

\date{\today}

\begin{abstract}
The Foucault Pendulum is a Spherical Pendulum of fixed length with two angular degrees of freedom,
attached to a suspension which rotates once a day around the Earth axis at a distance
essentially set by Earth radius and the geodetic latitude of the pendulum.
We write the Lagrange Function in the inertial frame of the fixed Earth axis and couple
it strictly to the rotating frame at which the suspension appears at rest. 
The Euler-Lagrange equations are
a coupled system of second-order differential equations for the two coordinates
of the mass
projected on the local horizontal plane. These are solved numerically in a C++ program
which allows to study the trajectories beyond the various standard approximations 
of the literature.
\end{abstract}

\maketitle

\section{Model of the Drag-Free Foucault Pendulum}
The mechanical model of the Foucault Pendulum in this work is a point mass $m$ fastened
to a suspension with a cord of length $l$ such that it follows the suspension---which
rotates at an angular speed $\Omega$ around the Earth axis---and follows a gravitational
force characterized by the acceleration $g$ in a homogeneous gravitational field.
The distance of the suspension to the Earth axis is essentially the product of Earth radius
and cosine of the geodetic latitude.
This concludes the set of free parameters in the model.
Wobbles of the Earth axis (polar motion or precession) are not included.

The master tool of this work is classical Lagrangian mechanics: the kinetic energy 
is a function of the squared time derivatives $(\dot X, \dot Y, \dot Z)$ of the mass position
$(X,Y,Z)$ in the inertial frame; the potential energy is a linear function of
the distance $z$ between the mass and the suspension in a local horizontal $(x,y,z)$ system
attached to the suspension. These two Cartesian coordinate systems are strictly
coupled by the circular motion at the point of suspension. The Euler-Lagrange differential equations for the two remaining
degrees of freedom (here chosen to be $x$ and $y$) are written down
without mentioning of pseudo-forces that are often introduced in standard texts
about the subject.

The fine art is that for example north-south excursions of the mass imply
small changes in its distance to the Earth axis which effect the precise
determination of centrifugal terms; also the speed along the east-west coordinate
is a local variation of the instantaneous angular speed that couples
to Coriolis forces \cite{NobleAJP20}.
This work handles these small effects by solving the
differential equations numerically, not entering discussions of which 
terms in the accelerations $\ddot x$ and/or $\ddot y$ represent named forces.

\section{Inertial versus Horizontal Coordinate Systems}
Let capital letters denote coordinates in some inertial 
frame with the Earth axis along the $+Z$ direction
and the $+X$ direction defined by a celestial reference
point like Aries.
In that non-rotating 
inertial Cartesian coordinate system attached to the Earth center 
the suspesion (subscript $s$)
moves as a function of time $t$ as
\cite[(3.152)]{RappGG1}
\begin{equation}
{\mathbf X}_s
=
\left(
\begin{array}{c}
X_s\\
Y_s\\
Z_s
\end{array}
\right)
=
\left(
\begin{array}{c}
(N+h)\cos\phi \cos(\Omega t)\\
(N+h)\cos\phi \sin(\Omega t)\\
{[N(1-e^2)+h]}\sin\phi 
\end{array}
\right)
\label{eq.Xs}
\end{equation}
where  \cite[(3.99)]{RappGG1}
\begin{equation}
N\equiv \frac{\rho_e}{\sqrt{1-e^2\sin^2\phi}}
\end{equation}
is the distance from the ellipsoid surface along the local vertical to the Earth axis,
where $\rho_e\approx 6378\times 10^3$ m is the equatorial radius \cite{NIMA8350},
$e\approx 0.081819$ the eccentricity of the Earth ellipsoid,
where $\phi$ is the geodetic latitude of the suspension, and
$h$ the altitude of the suspension above the ellipsoid \cite{NIMA8350,VermeilleJG76}.
The earth turns with an angular speed
of $\Omega\approx 7.2921\times 10^{-5}$ rad/s, $2\pi$ per sidereal day \cite{NIMA8350}.
The suspension moves on a circle of radius 
\begin{equation}
R\equiv (N+h)\cos\phi
\label{eq.Rdef}
\end{equation}
at constant angular velocity $\Omega$ around
the Earth axis.

The coordinates $x$, $y$ and $z$
centered at the suspension 
pointing east, north and up
span the horizontal coordinate system of the observer.
The unit vectors along the east, north and
up directions are given in the inertial system by
differentiating \eqref{eq.Xs} with respect to the geodetic longitude and latitude
The unit direction to the East is given in the inertial system by differentiating \eqref{eq.Xs}
with respect to the geodetic latitude $\Omega t$:
\begin{equation}
{\bf e}_E
=\left(
\begin{array}{c}
-\sin(\Omega t)\\
\cos(\Omega t)\\
0
\end{array}
\right);\quad
{\bf e}_E^2=1.
\end{equation}
The unit direction into the up direction is given in the inertial system by differentiating \eqref{eq.Xs}
with respect to the geodetic altitude $h$:
\begin{equation}
{\bf e}_U 
=\left(
\begin{array}{c}
\cos\phi \cos(\Omega t)\\
\cos\phi \sin(\Omega t)\\
\sin\phi
\end{array}
\right);\quad
{\bf e}_U^2=1; \quad {\mathbf e}_U\cdot {\mathbf e}_E=0.
\end{equation}
The unit direction to the North completes the right-handed orthogonal system:
\begin{equation}
{\bf e}_N
={\mathbf e}_U \times
{\mathbf e}_E
=\left(
\begin{array}{c}
-\sin\phi \cos(\Omega t)\\
-\sin\phi \sin(\Omega t)\\
\cos\phi
\end{array}
\right);\quad
{\bf e}_N^2=1;\quad
{\bf e}_N\cdot {\bf e}_E = {\bf e}_N\cdot {\bf e}_U=0.
\end{equation}
[A particularity of the geodetic coordinate system is that this is \emph{not} exactly the same
as the normalized derivative of \eqref{eq.Xs} with respect to $\phi$.]

The instantaneous location of the pendulum of mass $m$, cord length $l$ is
recorded by three Cartesian components in
that co-rotating horizontal (topocentric)
coordinate system \cite{MatharVixra1909}
\begin{equation}
{\mathbf x}=
\left(
\begin{array}{c}
x\\
y\\
z
\end{array}
\right)
=\left(
\begin{array}{c}
l\sin\varphi \cos \lambda\\
l\sin\varphi \sin \lambda\\
-l \cos\varphi
\end{array}
\right),
\label{eq.defAng}
\end{equation}
where $\varphi<90^\circ$ is the angle between cord and the vertical of the suspension, and $-180^\circ \le \lambda\le 180^\circ$
is an azimuth measured east to north.

The 
transformation from the topocentric horizontal to the global inertial
system is given by a matrix that contains the unit vectors 
$\mathbf e_{E,U,N}$ in the columns.
The location of the mass in the inertial frame is
\begin{equation}
\mathbf{X}
=
\mathbf{X_s}
+\mathbf{e}_E x
+\mathbf{e}_N y
+\mathbf{e}_U z
\label{eq.X}
\end{equation}
\begin{equation}
=
\left(
\begin{array}{c}
X\\
Y\\
Z\\
\end{array}
\right)
=
\left(
\begin{array}{c}
X_s\\
Y_s\\
Z_s
\end{array}
\right)
+
\left(
\begin{array}{ccc}
-\sin(\Omega t) & -\sin\phi\cos(\Omega t) & \cos\phi \cos(\Omega t)\\
\cos(\Omega t) & -\sin\phi\sin(\Omega t) & \cos\phi \sin(\Omega t)\\
0 & \cos\phi & \sin\phi
\end{array}
\right)
\cdot
\left(
\begin{array}{c}
x\\
y\\
z
\end{array}
\right)
\label{eq.Xofx}
\end{equation}

Time derivatives, the Cartesian components of the bob velocities in the horizontal coordinate system, are \cite{MatharVixra1909}
\begin{equation}
\dot{\mathbf x}
=
\left(
\begin{array}{c}
\dot x\\
\dot y\\
\dot z
\end{array}
\right)
=
l
\left(
\begin{array}{c}
 \dot \varphi \cos\lambda \cos \varphi - \dot \lambda \sin \varphi \sin \lambda ;\\
 \dot \varphi \sin\lambda \cos \varphi + \dot \lambda \sin \varphi \cos \lambda ;\\
 \dot \varphi \sin \varphi.
\end{array}
\right)
\end{equation}

In the horizontal frame the squared velocity is
\begin{equation}
\dot {\mathbf x}^2 = \dot x^2+\dot y^2+\dot z^2 = l^2(\dot \lambda^2\sin^2\varphi+\dot \varphi^2).
\end{equation}
and the positions and velocities of the bob are orthogonal:
\begin{equation}
{\mathbf x}\cdot \dot{\mathbf x}=0.
\label{eq.xxd}
\end{equation}
The Cartesian components of the velocity of the suspension in the inertial frame are
\begin{equation}
\dot {\mathbf X}_s
=
\left(
\begin{array}{c}
\dot X_s \\
\dot Y_s \\
\dot Z_s \\
\end{array}
\right)
=
(N+h)\Omega
\left(
\begin{array}{c}
- \cos\phi \sin(\Omega t)\\
\cos\phi \cos(\Omega t)\\
0
\end{array}
\right).
\end{equation}
The time derivative of \eqref{eq.X}, the velocity of the mass in the 
inertial coordinate system, is by the chain rule
\begin{equation}
\dot {\mathbf X}
=
\dot {\mathbf X}_s
+\dot {\mathbf e}_E x
+\mathbf{e}_E \dot x
+\dot {\mathbf e}_N y
+\mathbf{e}_N \dot y
+\dot {\mathbf e}_U z
+\mathbf{e}_U \dot z
\label{eq.Xdot}
.
\end{equation}
The time derivatives of the unit vectors of the local frame measured in the inertial frame are
\begin{equation}
\dot {\bf e}_E
=\Omega \left(
\begin{array}{c}
-\cos(\Omega t)\\
-\sin(\Omega t)\\
0
\end{array}
\right);\quad
\dot {\bf e}_N
=\Omega \left(
\begin{array}{c}
\sin\phi \sin(\Omega t)\\
-\sin\phi \cos(\Omega t)\\
0
\end{array}
\right);\quad
\dot {\bf e}_U 
=\Omega \left(
\begin{array}{c}
-\cos\phi \sin(\Omega t)\\
\cos\phi \cos(\Omega t)\\
0
\end{array}
\right).
\end{equation}

The square of \eqref{eq.Xdot}, basically the kinetic energy measured
in the inertial frame, is a second order polynomial of $\Omega$ with three
coefficients noted $\alpha_i(\varphi,\dot\varphi,\lambda,\dot \lambda)$
here:
\begin{equation}
\dot X ^2+\dot Y^2+\dot Z^2  \equiv \alpha_0 +\alpha_1\Omega +\alpha_2 \Omega^2
.
\end{equation}
With the aid of \eqref{eq.defAng} the right hand side can be rephrased 
with the angles $\varphi$, $\lambda$ and their
time derivatives
\begin{equation}
\alpha_0 = 
l^2\left( \dot \varphi^2 +\sin^2\varphi \dot \lambda ^2 \right)
;
\end{equation}
\begin{equation}
\alpha_1 
=
2(N+h)\cos\phi l\left(\dot \varphi \cos\varphi\cos\lambda - \dot \lambda \sin\varphi\sin\lambda\right)
+2l^2\left[
\sin\varphi\dot\lambda
(
\sin\phi \sin\varphi
+\cos\phi \cos\varphi\sin\lambda
)
-\cos\phi \dot\varphi \cos \lambda
\right]
;
\end{equation}
\begin{eqnarray}
\alpha_2 &=& 
\left[(N+h)\cos\phi
-l\left(\cos\phi\cos\varphi
+\sin\phi\sin\varphi\sin\lambda\right)\right]^2
+l^2\sin^2\varphi\cos^2\lambda
.
\end{eqnarray}

These terms have been published earlier: the variable substitution
$\dot l \to 0$, $\theta \to \pi-\varphi$, $\dot\theta \to -\dot\varphi$, $\phi\to \lambda-\pi/2$, $\dot\phi\to \dot\lambda$,
$\beta_{1,2}=0$ in \cite[(22)]{IcazaAM218} leads to the same notation.

Variants where terms $\propto \Omega^2$ are neglected have also been published 
\cite{BoulangerPhys2}\cite[p.\ 367]{JoseCM}.

The key point of this approach is to bypass any ad-hoc insertions of Coriolis and centrifugal terms
which arise in other publications on the Foucault Pendulum \cite{SomervilleQJRAS13}.

The intent is to predict long-time motions of the Foucault Pendulum 
accurately in the drag-free limit \cite{CartmellPRSA2019,CartmellNCPS2021,PippardPRSL420}.

\section{Lagrangian}
Kinetic energy $K$ and potential energy $V$ of the mass are
\begin{equation}
K=\frac12 m (\dot X^2+\dot Y^2+\dot Z^2) = \frac12 m (\alpha_0 +\alpha_1\Omega +\alpha_2\Omega^2);
\label{eq.K}
\end{equation}
\begin{equation}
V^{(h)}=mgz = -mgl \cos\varphi.
\end{equation}
There is no need to mix terms of $K$ and $V$ for some sort of apparent 
gravity \cite{PerssonQJRMS141,PhillipsBAMS81,DurranAMS74}.

In the limit $\Omega\to 0$ these two differential equations reduce to those of the spherical pendulum.

Lagrangian mechanics is valid because 
the constraints of the positions,
which means the sum of the $x^2+y^2+z^2=l^2$,
 are holonomic.

Since $V$ does not depend on $\lambda, \dot\lambda$, 
\begin{multline}
\partial {\cal L}/\partial \lambda
=
\partial K/\partial \lambda
=
\frac12 m(
\partial \alpha_0/\partial \lambda
+\partial \alpha_1/\partial \lambda \Omega
+\partial \alpha_2/\partial \lambda \Omega^2
)
\\
=
\frac12 m(
[ 2(N+h)\cos\phi l\left(-\dot\varphi \cos\varphi\sin\lambda-\dot\lambda\sin\varphi\cos\lambda\right) 
+2l^2\left(\sin\varphi\dot\lambda\cos\phi\cos\varphi\cos\lambda+\cos\phi\dot\varphi\sin\lambda\right)
]\Omega
\\
+[
-2l[(N+h)\cos\phi-l(\cos\phi\cos\varphi+\sin\phi\sin\varphi\sin\lambda)]
\sin\phi\sin\varphi\cos\lambda
-2l^2\sin^2\varphi\cos\lambda\sin\lambda
]\Omega^2
)
.
\end{multline}
The Euler-Lagrange equation
\begin{equation}
\frac{d}{dt}\frac{\partial {\cal L}}{\partial \dot\varphi}
-\frac{\partial {\cal L}}{\partial \varphi}=0
\end{equation}
expands to

\begin{multline}
\ddot \varphi
+\hat R \Omega^2 (\sin\phi \cos\varphi\sin\lambda
-\cos\phi \sin\varphi)
+2\Omega\sin\varphi( \cos\phi \sin\varphi \sin\lambda 
-
\sin\phi \cos\varphi
)\dot\lambda
\\
+\Omega^2 (\sin\phi\sin\varphi+\cos\phi\cos\varphi\sin\lambda)(\cos\phi \sin\varphi \sin\lambda -\sin\phi \cos\varphi
)
\\
-\cos\varphi\sin\varphi\dot\lambda^2
+\frac{g}{l}\sin\varphi=0.
\label{eq.pdd}
\end{multline}

\begin{equation}
\frac{d}{dt}\frac{\partial {\cal L}}{\partial \dot\lambda}
-\frac{\partial {\cal L}}{\partial \lambda}=0
\end{equation}
leads to
\begin{equation}
\sin\varphi \ddot \lambda
+2\cos\varphi \dot \varphi \dot\lambda
+\hat R \Omega^2 \sin\phi \cos\lambda
+\Omega (\cos\phi \sin\varphi \sin\lambda -\sin\phi \cos\varphi)
(\Omega\cos\phi\cos\lambda -2\dot\varphi)
=0.
\label{eq.ldd}
\end{equation}
The notation with a roof top over a quantity indicates division
through the cord length:
\begin{equation}
\hat R \equiv (N+h)\cos\phi/l
\end{equation}
derived from \eqref{eq.Rdef}.

The Lagrangian does not explicitly depend on time,
but the energy $E=K+V$ measured in the inertial frame is \emph{not} conserved \cite{DeslogeAJP45}.

If the three terms of the kinetic energy \eqref{eq.K} are
not written in terms of the angles $\varphi$ and $\lambda$
but in terms of the Cartesian $x$ and $y$ coordinates,
the expansion coefficients are
\begin{equation}
\alpha_0 = \frac{2xy\dot x\dot y+(l^2-y^2)\dot x^2+(l^2-x^2)\dot y^2}{l^2-x^2-y^2};
\end{equation}
\begin{multline}
\alpha_1 = [-2\dot x \sin\phi y l^2
-2\dot x \cos\phi (l^2-x^2-y^2)^{3/2}
+2\dot x \sin\phi y^3
-2\dot y \sin\phi x^3
+2\dot x \sin\phi yx^2
\\
-2\dot x \cos\phi x^2(l^2-x^2-y^2)^{1/2}
-2\dot y \cos\phi xy(l^2-x^2-y^2)^{1/2}
+2l^2(N+h)\cos\phi \dot x
-2(N+h)\cos\phi x^2\dot x 
\\
-2(N+h)\cos\phi y^2\dot x 
+2l^2\sin\phi x\dot y 
-2\sin\phi xy^2\dot y 
] /(l^2-x^2-y^2);
\label{eq.alph1xy}
\end{multline}
\begin{multline}
\alpha_2 = [
-2x^2y^2
+y^2l^2
+2\cos^2\phi y^4+\cos^2\phi x^4
+x^2l^2+\cos^2\phi l^4-x^4-y^4
+(N+h)^2\cos^2\phi l^2
-3y^2l^2\cos^2\phi
\\
-2(N+h)\cos^2\phi(l^2-x^2-y^2)^{3/2}
-2\cos^2\phi l^2 x^2
+3\cos^2\phi x^2y^2
+2\sin\phi \cos\phi y (l^2-x^2-y^2)^{3/2}
\\
-(N+h)^2\cos^2\phi x^2
-(N+h)^2\cos^2\phi y^2
-2(N+h)\sin\phi \cos\phi yl^2
\\
+2\sin\phi\cos\phi (N+h)yx^2
+2\sin\phi\cos\phi(N+h)y^3
] /(l^2-x^2-y^2)
.
\label{eq.alph2xy}
\end{multline}
[The main reason to switch from angular to Cartesian coordinates is that the sine factor
in the first term of \eqref{eq.ldd} induces quick changes of $\lambda$ if the pendulum approaches
small $\varphi$. This requires some type of step size control in numerical integrations
which is unlikely needed for $x$ and $y$ which are almost sinusoidal in time.]

\section{In Cartesian Coordinates}
The standard steps that follow are writing down the two Lagrange equations
starting with
\begin{equation}
\frac{d}{dt}\frac{\partial {\cal L}}{\partial \dot x}
-\frac{\partial {\cal L}}{\partial x}=0.
\end{equation}
After having evaluated all time derivatives, we may multiply
the result with $z^2$ to have less cluttered denominators
(assuming $z\neq 0$, the mass staying below the suspension), to yield
\begin{equation}
(l^4+y^4-2y^2l^2
+y^2x^2-l^2x^2
)
\ddot x
+(-x^3y+xyl^2-xy^3)\ddot y
+\sum_{i=0}^2 \beta_{x,i}\Omega^i=0.
\end{equation}
For compact notation the orders of $\Omega$ are bundled in
three $\beta_{x,i}$ coefficients

\begin{equation}
\beta_{x,0} 
=
x[
l^2(\dot x^2+\dot y^2)
-(x\dot y-y\dot x)^2
]
+xg
|z|^3
;
\label{eq.betax0}
\end{equation}
\begin{equation}
\beta_{x,1}
=
-2z^4
\sin\phi \dot y
+2y\dot y\cos\phi |z|^3
;
\label{eq.betax1}
\end{equation}

\begin{equation}
\beta_{x,2}
=
-x z^4
\sin^2\phi 
+x\cos\phi
(y\sin\phi -R)|z|^3
;
\label{eq.betax2}
\end{equation}
\begin{equation}
z^2(x^2+z^2)
\ddot x
+z^2yx\ddot y
+\sum_{i=0}^2 \beta_{x,i}\Omega^i=0
.
\label{eq.betax}
\end{equation}
Equivalent computation for
\begin{equation}
\frac{d}{dt}\frac{\partial {\cal L}}{\partial \dot y}
-\frac{\partial {\cal L}}{\partial y}=0
\end{equation}
expands to
\begin{equation}
z^2yx\ddot x
+z^2(y^2+z^2) \ddot y
+\sum_{i=0}^2 \beta_{y,i}\Omega^i=0.
\label{eq.betay}
\end{equation}
\begin{equation}
\beta_{y,0} 
=
y[
l^2(\dot x^2+\dot y^2)
-(x\dot y-y\dot x)^2
]
+yg
|z|^3;
\label{eq.betay0}
\end{equation}
This is obtained from $\beta_{x,0}$ by flipping the roles of $x$ and $y$.
The symmetry argument is that in the limit $\Omega \to 0$ there is no
bias in the oscillations with respect to the compass directions.

\begin{equation}
\beta_{y,1}
=
2z^4
\sin\phi \dot x
-2y\dot x\cos\phi |z|^3
;
\label{eq.betay1}
\end{equation}
This is obtained from $\beta_{x,1}$ by flipping the roles of $x$ and $y$ and switching the sign.

\begin{equation}
\beta_{y,2}
=
z^4(
-y + R\sin\phi + 2 y\cos^2\phi
)
-|z|^3\cos\phi
[Ry+(z^2-y^2)\sin\phi]
;
\label{eq.betay2}
\end{equation}
The matrix-vector form of the differential equations \eqref{eq.betax} and \eqref{eq.betay} 
is
\begin{equation}
\left(\begin{array}{cc}
z^2(x^2+z^2) & z^2yx \\
z^2yx & z^2(y^2+z^2 \\
\end{array}
\right)
\cdot
\left(\begin{array}{c}
\ddot x\\
\ddot y\\
\end{array}
\right)
=
\left(\begin{array}{c}
-\sum_{i=0}^2 \beta_{x,i}\Omega ^i\\
-\sum_{i=0}^2 \beta_{y,i}\Omega ^i
\end{array}
\right)
.
\end{equation}
To decouple them 
we multiply by the inverse of the $2\times 2$ matrix:
\begin{equation}
\left(\begin{array}{c}
\ddot x\\
\ddot y\\
\end{array}
\right)
=
\frac{1}{z^4l^2}
\left(\begin{array}{cc}
y^2+z^2 & -yx \\
-yx & x^2+z^2 \\
\end{array}
\right)
\cdot
\left(\begin{array}{c}
-\sum_{i=0}^2 \beta_{x,i}\Omega ^i\\
-\sum_{i=0}^2 \beta_{y,i}\Omega ^i
\end{array}
\right)
\label{eq.main}
\end{equation}
such that the second derivatives depend only
on the positions and first derivatives, an initial value problem.

\section{Stable Kinematic Position}\label{sec.rest}
At which point in the local frame does the pendulum
obtain a stable, zero-velocity position?
Where do the forces on the pendulum mass (gravitational, cord
and forces implied by the motion within the inertial frame)
cancel such that the mass ``rests'' in the horizontal ``rest'' frame tied to the Earth crust?
(This question arose during commissioning
of the SDSS-V LVM telescope benches where someone proposed
to measure the local south direction by looking at a plumb line's shade
in the sun at some specific time. The aim was to avoid ambiguities from
wandering of the magnetic poles. This is the historical impetus for writing this 
manuscript.)

We ask for the position where the $x$ and $y$ coordinates
remain frozen in the local reference frame, i.e., 
where $\ddot x=\ddot y =\dot x=\dot y=0$,
and in consequence $\beta_{x,1}=\beta_{y,1}=0$.
Solving \eqref{eq.main} in this case yields $x=0$, because $\beta_{x,0}$ and $\beta_{x,1}$ are
both proportional to $x$ and because the upper right element of the matrix is also proportional to $x$.
So that position is on the meridian running through the suspension.

This satisfies the upper equation; the lower then requires
\begin{equation}
0= \frac{1}{z^4l^2}z^2(-\beta_{y,0}-\beta_{y,2}\Omega^2),
\end{equation}
which is a quartic equation for $y$:
\begin{equation}
y^2= \sqrt{l^2-y^2}(y-R\sin\phi-2y\cos^2\phi)\frac{\Omega^2}{g} +\cos\phi\left[Ry+(l^2-2y^2)\sin\phi\right]\frac{\Omega^2}{g}.
\end{equation}
For $l$ in the range 1 to 100 m this produces Figure \ref{fig.excur}; 
the typical deviations from the vertical are less than $0.1^\circ$.
(Local microgravity is obviously neglected.)
There are two derivations of the approximation, an arithmetic and 
a physical:
\begin{enumerate}
\item
It is easier to solve the problem in the polar 
angular
$(\varphi,\lambda)$ 
variables than in the Cartesian $(x,y)$ variables. We know already
that the position is on the meridian, so $\lambda =\pm 90^\circ$,
$\cos\lambda=0$. The requirement of rest means 
$\ddot \lambda=\dot \lambda=\dot \varphi=0$
so all terms in \eqref{eq.ldd} vanish; the equation is satisfied.
In \eqref{eq.pdd} we know $\sin\lambda=\pm 1$, $\dot\lambda=0$, so
\begin{multline}
\hat R \Omega^2 (\pm \sin\phi \cos\varphi
-\cos\phi \sin\varphi)
\\
+\Omega^2 (\sin\phi\sin\varphi\pm\cos\phi\cos\varphi)(\pm\cos\phi \sin\varphi -\sin\phi \cos\varphi
)
+\frac{g}{l}\sin\varphi
\\
=
\pm \hat R \Omega^2 \sin(\phi\mp \varphi)
\mp \Omega^2 \cos(\phi\mp \varphi) \sin(\phi\mp \varphi)
+\frac{g}{l}\sin\varphi
=0
.
\end{multline}

Series expansion in orders of $\Omega^2$ and reversion of that series 
up to $O(\Omega^6)$
yields
\begin{multline}
\varphi \approx 
\mp \sin(2\phi)(N+h-l)\Big\{\frac{1}{2g}\Omega^2
+ \frac{(N+h)\frac{1+\cos(2\phi)}{2}-l\cos(2\phi)}{2g^2}\Omega^4
\\
+ \frac{(N+h)^2[\cos(4\phi)+3\cos(2\phi)+2]-(N+h)l[\frac{37}{8}\cos(4\phi)+\frac{11}{8}+6\cos(2\phi)]+l^2[\frac{35}{8}\cos(4\phi)+\frac{13}{8}])}{12g^3}\Omega^6
+O(\Omega^8)
\Big\}.
\label{eq.centr}
\end{multline}
$\varphi$ is always positive and the upper/lower
sign of $\lambda$ must ensure this for both signs of $\phi$,
so the sign of $\lambda$ is the opposite sign of $\phi$.
\begin{figure}
\includegraphics[scale=0.7]{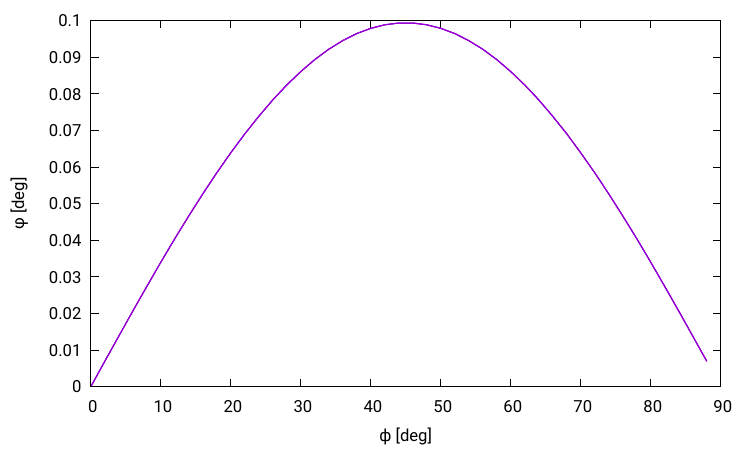}
\caption{
Angular deviation $\varphi$ of the rest position from the local geodetic vertical 
for pendulums at geodetic latitudes $\phi$ on Earth.
These are three curves for $l=1, 10$ and $100$ m which are indistinguishable
on this scale.
}
\label{fig.excur}
\end{figure}

\item
A physicist draws  the force triangle 
of Figure \ref{fig.centrif} in which the acceleration $g$ points
towards the earth center, the centrifugal acceleration $R\Omega^2$ points at an angle $\phi$ relative
to it, and where $\varphi$ is a small angle in that triangle. 
\begin{figure}
\includegraphics[scale=0.5]{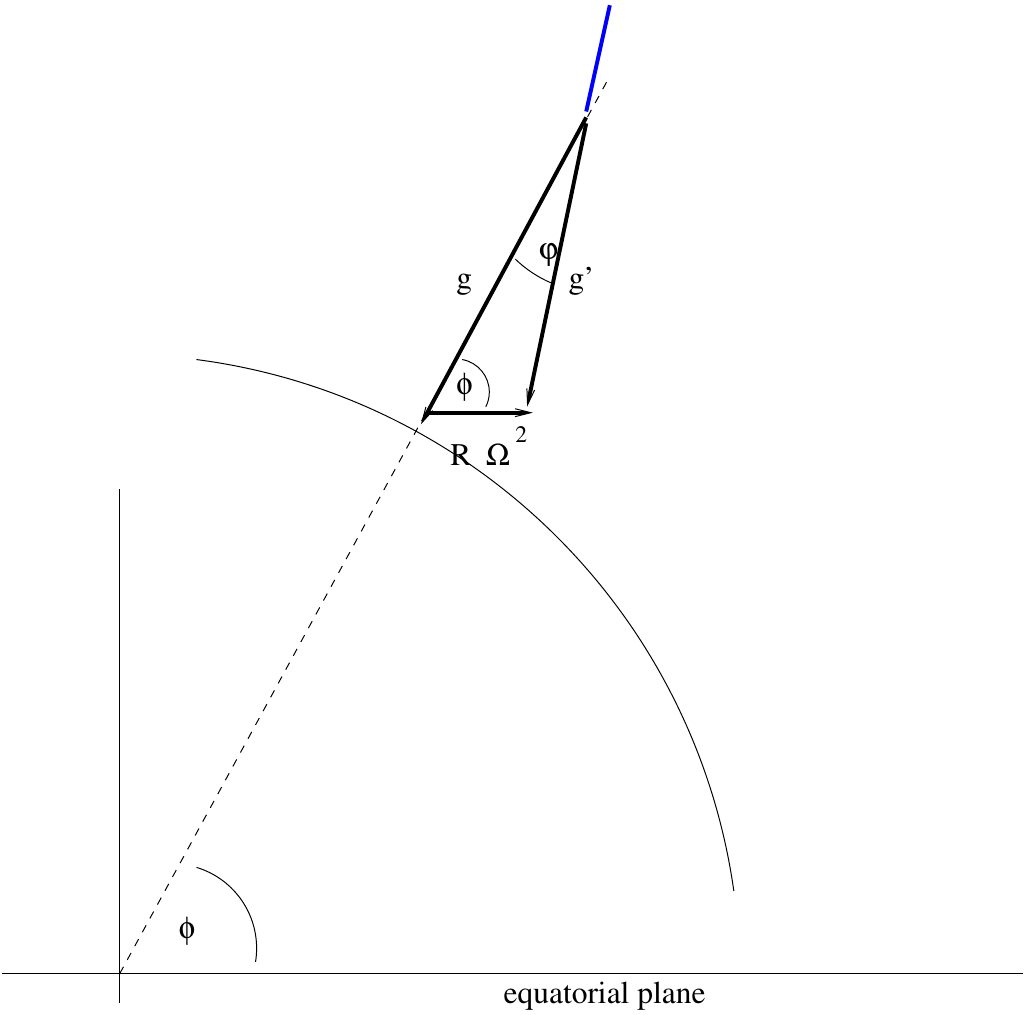}
\caption{Triangle of forces in a cross section through the Earth with the pendulum at geographic latitude $\phi$: 
gravitational acceleration $g$, centrifugal acceleration $R\Omega^2$ parallel to the equatorial plane,
their vector sum $g'$,
and the direction (blue) of the rod parallel to $g'$.
$\varphi$ is the mismatch between
the local geodetic down direction and the kinematically balanced direction of a plumb line.
}
\label{fig.centrif}
\end{figure}
Then the third side
in that triangle (the acceleration to be compensated by the pendulum) is by the law
of cosines 
\begin{equation}
g'=\sqrt{g^2+(R\Omega^2)^2-2gR\Omega^2\cos\phi}
.
\label{eq.gcentr}
\end{equation}
The desired
angle $\varphi$ is extracted with the sine rule 
\begin{equation}
\frac{R\Omega^2}{\sin\varphi}=\frac{g'}{\sin\phi};
\end{equation}
\begin{equation}
\leadsto \sin\varphi 
= \frac{R\Omega^2 \sin\phi}{\sqrt{g^2+(R\Omega^2)^2-2gR\Omega^2\cos\phi}}
\end{equation}
and a Taylor expansion for small $R\Omega^2=(N+h)\cos\phi \Omega^2$ is
$$
\sin\varphi \approx \frac{\sin\phi}{g}R\Omega^2+\frac{\cos\phi\sin\phi}{g^2}(R\Omega^2)^2+O((R\Omega^2)^3)
\approx \frac12 \frac{\sin(2\phi)}{g}(N+h)\Omega^2+O((R\Omega^2)^2)
$$
compatible with \eqref{eq.centr}
and footnote 5 in
\cite{KrivoruPU52}.
\end{enumerate}

\section{Weak Coupling / Slow rotation}\label{sec.approx}
The standard harmonic approximation for the $x-y$ motion 
is derived from the full-fledged
\eqref{eq.main}
by the following
approximations:
\begin{enumerate}
\item
The terms proportional to $\Omega^2$ are neglected with the argument that this is of the 
order of $5\times 10^{-9}$ numerically and/or that these terms stem from effects that would
persist if the direction of the Earth Rotation axis would be flipped. Effectively
this replaces $\beta_{x,2}=\beta_{y,2}\to 0$
\cite{MacMillanAJM37}.

A variant of this approximation actually keeps the terms $\propto R\cos\phi$ in \eqref{eq.betax2}
and \eqref{eq.betay2} and continues by using the gravitational
acceleration  $\bar g \equiv g-R\Omega^2\cos\phi$ in the sequel
\cite{PhillipsBAMS81}. [This is the first order Taylor expansion of \eqref{eq.gcentr} for small $R\Omega^2$.]
\item
The outer diagonal terms $-yx$ in the $2\times 2$ matrix in \eqref{eq.main} are set to zero arguing that
only oscillations of small $x$ and $y$ are treated, actually
smaller than $z^2\sim l^2$ in the diagonal terms.
\item
In \eqref{eq.betax0} and \eqref{eq.betay0} only the terms that contain $g$ are kept, again with
the small oscillations argument, assuming that the terms with the time derivatives are smaller
than terms of the associated orders of the pendulum length.
\item
In \eqref{eq.betax1} and \eqref{eq.betay1} only the terms with $\sin\phi$ but not
the terms with $\cos\phi$ are kept, again with
the small oscillations argument that one power of $z$ outweighs a power of $y$ or $x$.
\item
$z$ is replaced by $l$ with the argument that its relative changes are not important.
\end{enumerate}
This approximation of \eqref{eq.main} is \cite{SchulzAJP38,DasCJP52,ConduracheIJNLM43,BromwichPLMS13,OpatAJP59}
\begin{eqnarray}
\hat{\ddot x} &=& -\hat x\hat g +2\Omega \sin\phi \hat{\dot y}; \label{eq.xstd}\\
\hat{\ddot y} &=& -\hat y\hat g -2\Omega \sin\phi \hat{\dot x} \label{eq.ystd}.
\end{eqnarray}
with 
$\hat{\ddot x}\equiv \ddot x/l$, $\hat{\ddot y}\equiv \ddot y/l$,
$\hat{\dot x}\equiv \dot x/l$, $\hat{\dot y}\equiv \dot y/l$,
$\hat g\equiv g/l$.

\section{Example: 67 m Pendulum in Paris}
The properties of the trajectory are demonstrated with parameters of a pendulum
of $l=67$ m of length at a geodetic latitude $\phi=48.846111^\circ$, referred
to as the ``Paris pendulum'' in the sequel
\cite{SommeriaCRP18}.

\begin{figure}
\includegraphics[scale=0.7]{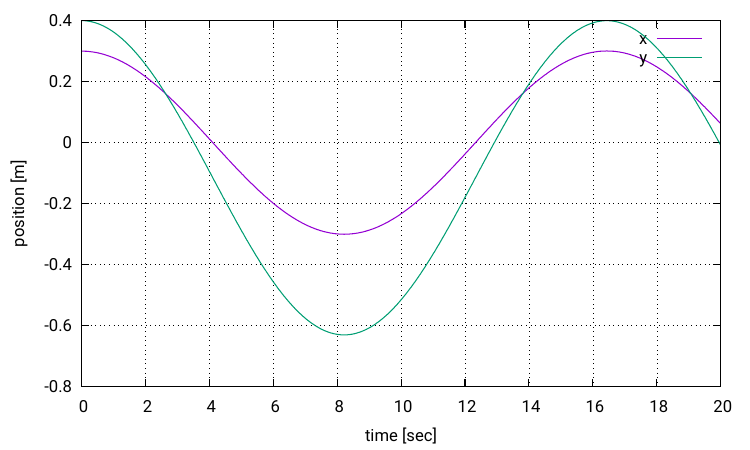}
\includegraphics[scale=0.7]{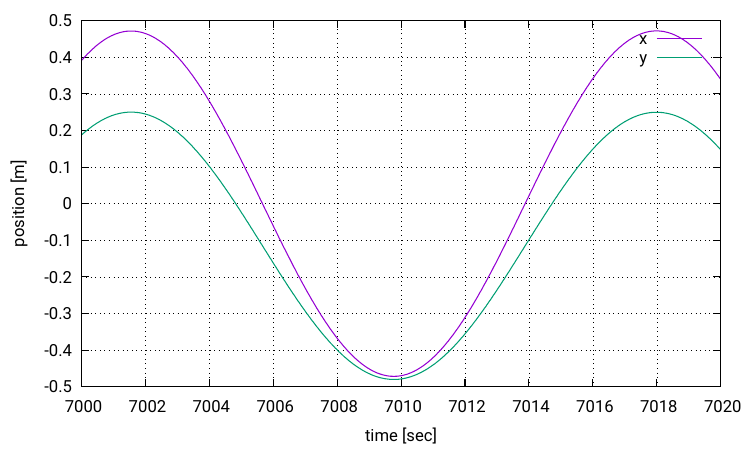}
\includegraphics[scale=0.7]{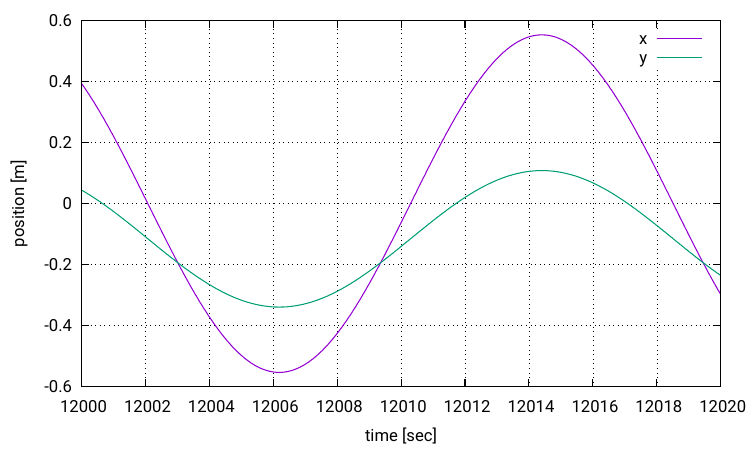}
\caption{
Initial 20 seconds of the trajectory in $x$ and $y$ for the Paris pendulum released at $x=0.3$ m
and $y=0.4$ m, another interval of 20 seconds after 7000 seconds have passed, and another  20 seconds when 12000 seconds
have passed.
}
\label{fig.xyParis}
\end{figure}

\begin{figure}
\includegraphics[scale=0.8]{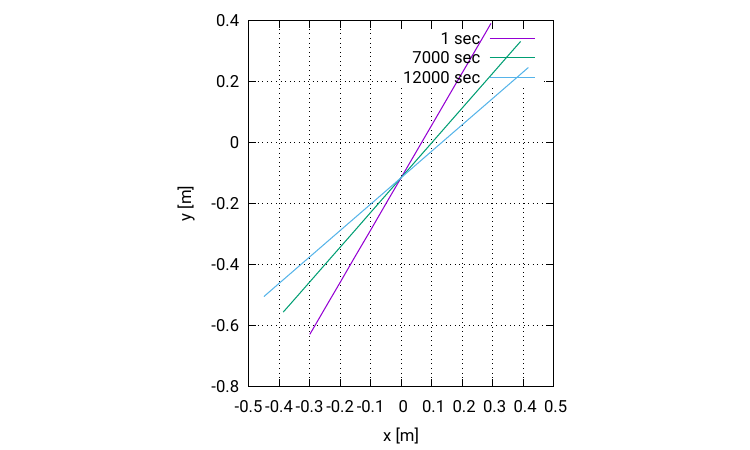}
\includegraphics[scale=0.8]{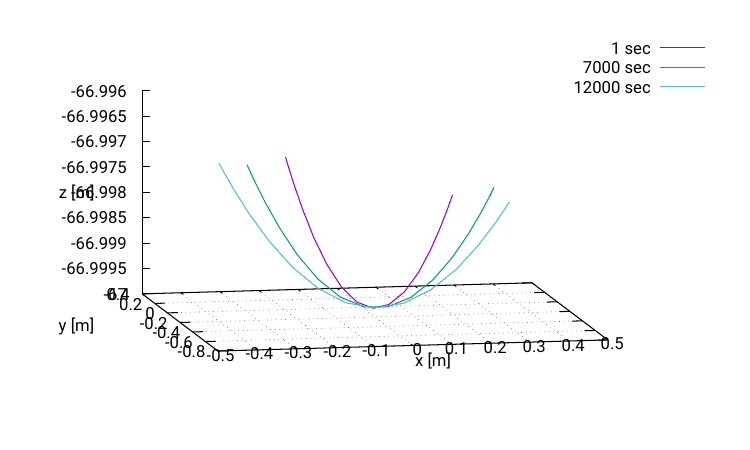}
\caption{
Three trajectories each
of 17 seconds of duration equivalent to snapshots of Figure \ref{fig.xyParis}
in a 3D view and in the projection on the $x-y$-plane, demonstrating the slow rotation
of the main plane of the pendulum motion around the point equivalent to the rest position.
}
\label{fig.xyzParis}
\end{figure}

It is released with zero velocity
in the rest frame at time zero at $(x,y)=(0.3,0.4)$ m, displaced 30 cm to the East
and 40 cm to the North away from the point under the suspension, which implies roughly 2 mm above the point
of lowest gravitational potential. Figures \ref{fig.xyParis} and
\ref{fig.xyzParis} illustrate the time evolution of the trajectories.
The first swing leads to the puzzling consequence to end
at a few millimeters higher $z$-value than where the pendulum
was released. The projections in Figure \ref{fig.xyzParis} illustrate
that the actual center of the motion is at the point  further south
characterized in Section \ref{sec.rest} as the kinematic point of rest.
The centrifugal force manages to inject energy to create
such an apparent violation of the principle of energy conservation.
As stated earlier \cite{IcazaAM218,ChessinAJM17}, the pendulum never 
moves through the vertical at $x=y=\varphi=0$ if started with
zero velocity in the local frame.

\begin{figure}
\includegraphics[scale=0.8]{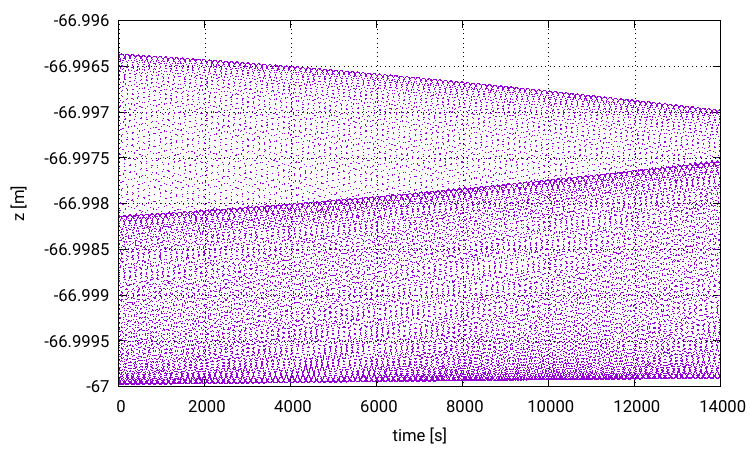}
\caption{Snapshots of the $z$-coordinate of the Paris pendulum.
}
\label{fig.z}
\end{figure}

\begin{figure}
\includegraphics[scale=0.8]{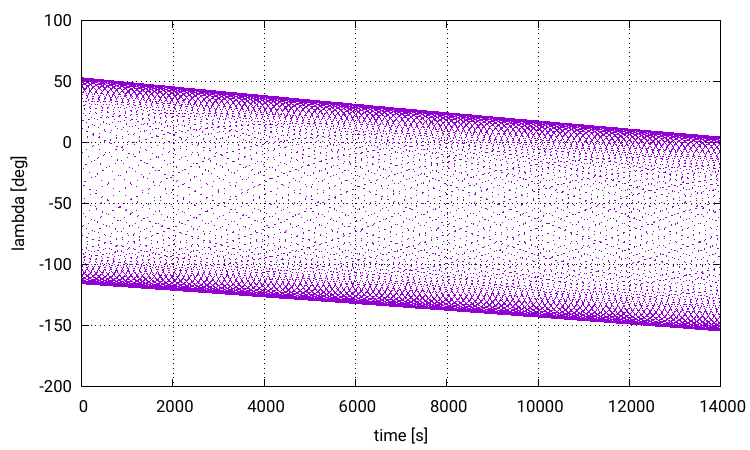}
\caption{Snapshots of the $\lambda$-angle of the Paris pendulum.
}
\label{fig.lambd}
\end{figure}

Figure \ref{fig.z} are snapshots of the $z$ coordinates
of the pendulum at regular time intervals. 
The points of small velocity at the turning points create two denser
clouds of points; one is changing
 from the point of release near $-66.98$ upwards in time, the other from
the opposite points near $-66.9965$ downwards in time.
The slow rotation from a NE-SW to more E-W
orientation of the mean plane of motion removes the obvious bias
in the altitudes of the early periods.
Figure \ref{fig.lambd} illustrates that the rotation of the
mean azimuth $\lambda$ over time is very linear.

\begin{figure}
\includegraphics[scale=0.8]{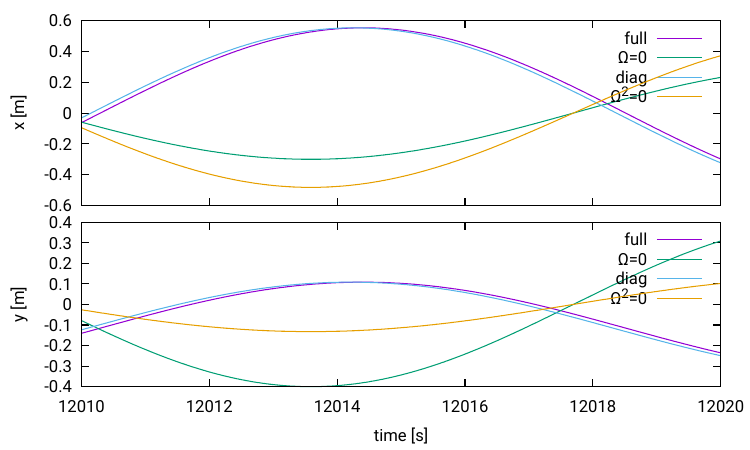}
\caption{The difference in positions between the full theory and 
different approximations introduced in Section \ref{sec.approx}.
}
\label{fig.ta1}
\end{figure}

Figure \ref{fig.ta1} shows the trajectories of the Paris pendulum
after 12 thousand seconds. The green lines are reference positions
obtained by switching the Foucault effect off, i.e., by setting $\Omega=0$
and looking at the motions of the planar pendulum in the range $-0.3<x<0.3$ and $-0.4<y<0.4$.
The magenta curves are the positions of the full theory; the blue
curves with the approximation of ignoring
the outer-diagonal terms as in item 2 in Section \ref{sec.approx}
are a little bit ahead in time.
The orange curves are created by dropping terms of order $\Omega^2$
as in item 1 in Section \ref{sec.approx}.
The Foucault pendulum has a frequency $\approx 0.06084$ Hz, and the planar pendulum 
a frequency $\approx 0.06089$ Hz. 
After 12 thousand seconds the two pendulums are detuned by approximately half a cycle,
i.e., the green curve is near a minimum where the blue/magenta curves are near a maximum.

There are various approximations to the frequency of the oscillations:
\begin{itemize}
\item
The harmonic (small-angle) approximation of the planar pendulum has an angular frequency $\sqrt{g/l}$,
which (after division through $2\pi$) is a frequency of $0.0608895$ Hz.
\item
The anharmonic planar pendulum released at a projected distance of $\sqrt{0.3^2+0.4^2}$ m from 
its rest position equivalent to $\varphi^{(0)}\approx 0.00746276$ rad has a larger period given by 
$4\sqrt{l/g}K(\sin(\varphi^{0}/2))\approx 16.423225$ seconds, which is $0.06088938$ Hz.
Here $K$ is the Complete Elliptic Integral of the First Kind, not the kinetic energy.
\item
The circular frequency deduced from \eqref{eq.xstd}--\eqref{eq.ystd} is
$\sqrt{\hat g+(\Omega \sin\phi)^2}$ \cite{BromwichPLMS13}
equivalent to a frequency $\approx 0.06088959$ Hz.
\item
The same as in the previous bullet 
but replacing $g=9.80665$ m/s$^2$ by 
the ``observed'' acceleration of \eqref{eq.gcentr},
$g'=9.79195$ m/s$^2$. 
This frequency is $\approx 0.060843935$ Hz.
\item
The same as in the previous bullet 
plus a slow-down with the factor $4K(\sin(\varphi^0/2))$ for the anharmonic oscillator, where
the angular amplitude $\varphi^{(0)}$ is taken as the distance from the point of release
at $(x,y,z)=(0.3,0.4,-66.9981)$ m to the ``equilibrium'' center of $(x,y,z)=(0,-0.115202,-66.999)$ m
obtained in Section \ref{sec.rest}, $\varphi^{(0)}=0.008898355$ rad. This yields 0.060843634 Hz.
\item
The actual Foucault pendulum traced  by numerical integration
of the coupled differential equations at a frequency of 0.060843633 Hz. This is one of the data points in Figure \ref{fig.foflat}.
\begin{figure}
\includegraphics[scale=0.8]{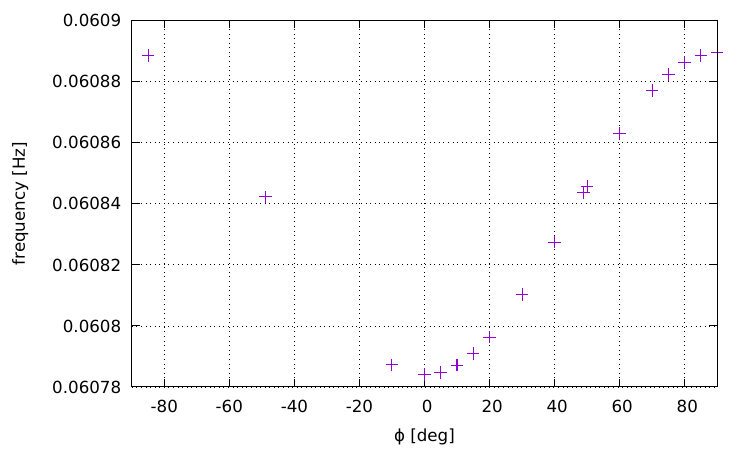}
\caption{The frequency of the pendulum as a function of the
geodetic latitude $\phi$, measured by time differences between transits $\dot x=0$.
}
\label{fig.foflat}
\end{figure}
\end{itemize}

\section{Summary}
We have written down the coupled second-order differential equations for
the East-West and North-South components of the projected Cartesian horizontal coordinates
of the Foucault Pendulum derived from a strict treatment of the Lagrangian.
This is implemented in a C++ program with variable Runge-Kutta-Nystr\"om orders
of the step control for the initial value problem.

An accurate estimate of the frequency of the pendulum follows from matching
the motion with a (slightly anharmonic) plane pendulum that moves through
the ``equilibrium'' point, as defined by the position where gravity
and centrifugal force are balanced by the rod.

\appendix 
\section{C++ Implementation}
A numerical integration of the system \eqref{eq.main} of two differential equations
is implemented in the C++ program in the \texttt{anc} directory. It is compiled by
calling a C++ compiler (as in the \texttt{Makefile}) that creates the binary \texttt{foucPend}.
The main program is called with the following command line options:

\texttt{foucPend [-x} \textit{xCartes}\texttt{] [-y } \textit{yCartes}\texttt{] [-l }\textit{cordlength}
\texttt{] [-v} \textit{veloc}
\texttt{] [-L} \textit{geolat}
\texttt{] [-T} \textit{timedurat}
\texttt{] [-t} \textit{timestep}
\texttt{] [-s} \textit{sampleskip}
\texttt{] [-g} \textit{gconstant}
\texttt{] [-O} \textit{Omega}
\texttt{] [-f} \textit{invflat}
\texttt{] [-h} \textit{geoalt}
\texttt{] [-r} \textit{earthMajRad}
\texttt{] [-R} \textit{RKorder}
\texttt{] [-a} \textit{approxFlags}
\texttt{]}
The square brackets indicate optional parameters and are not part of the syntax.
The meaning and defaults of the parameters are
\begin{itemize}
\item
\texttt{-x} horizontal $x$-component of the position at time zero in meters, east positive. Zero if not used.
\item
\texttt{-y} horizontal $y$-component of the position at time zero in meters, north positive. Zero if not used.
\item
\texttt{-l} cord length in meters. 67 if not used.
\item
\texttt{-v} equatorial component of the velocity in the local rest frame at time zero, meters per second. Positive if 
   starting in the direction according to the right-hand-rule relative to the spin of $\Omega$. Zero if not used.
\item
\texttt{-L} geodetic latitude $\phi$ in degrees. 48.846111 if not used.
\item 
\texttt{-T} duration of the trajectory to be integrated, in seconds. Default is $24\times 3600$,
that means approximately 4 minutes longer than a sidereal day.
\item 
\texttt{-t} Time steps in the Runge-Kutta integration scheme. Default is 0.1 seconds.
\item 
\texttt{-s} Subsampled time steps before one line of output is created, $\ge 1$. Which means 
    after each $s$th step in the numerical solution of the differential equations one 
    snapshot of the trajectory is printed. The output of the program shrinks by that factor without
    reducing the numerical precision (the latter set by the \texttt{-t} and \texttt{-R} switches).
\item 
\texttt{-g} Local constant of gravitational acceleration in meters per second squared. Default is 9.80665.
\item 
\texttt{-O} The angular speed of the earth axis in radians per second. The default is $7.292115\times 10^{-5}$.
\item 
\texttt{-f} The inverse flattening assumed in the prolate elliptical coordinates. The default is 298.257223564.
\item 
\texttt{-h} Geodetic altitude of the pendulum suspension above the reference ellipsoid in meters. The default is zero.
\item 
\texttt{-r} The equatorial radius of the earth in meters. The default is $6.378137\times 10^6$.
\item 
\texttt{-R} a value from 4 to 6 of the numerical integration scheme. The default is 4.
  A value of 4 is a classical 4-step Runge-Kutta integration \cite[p. 242]{EngelnMullges}\cite{RutishauserNM2}.
  (The alternative proposed in \cite{FehlbergZAMM67} is not implemented.)
  A value of 5 is the RKN-G-5(6)-8 integration as published in \cite{FehlbergTRR432,FehlbergComp14}.
  A value of 6 is the RKN-G-6(7)-10 integration as published in \cite{FehlbergTRR432,FehlbergComp14}.
  The latter two are Runke-Kutta-Nystr\"om schemes with order 6 (resp 7.) for generalized (here: second order)
  systems of differential equations with 8 (resp. 10) stages of evaluating the right hand side
  of \eqref{eq.main} for each step. Step size control is not implemented; it is the constant specified
  by the \texttt{-t} option. The kind of numerical error for a \texttt{-t} selection of 0.5 seconds
  is illustrated in Figure \ref{fig.R}.

\begin{figure}
\includegraphics[scale=0.9]{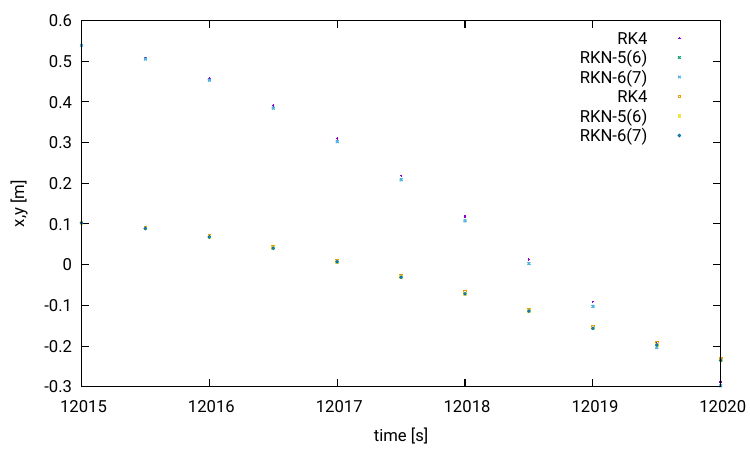}
\caption{
Illustration of the effect of the 3 different Runge-Kutta-Nystr\"om order parameters 
through the \texttt{-R} option of the program
on the Paris pendulum positions at a step size of $0.5$ seconds after 12000 seconds
have passed. (The main results of this paper have been obtained with step sizes of 0.01 seconds.)
}
\label{fig.R}
\end{figure}
\item 
\texttt{-a} specifies an integer which is interpreted bitwise as a non-negative binary number.
  The default is 0, which means the full theory with all terms of the equations in this paper is employed.
  The implemented approximations are:
  \begin{enumerate}
  \item
  If the least-significant bit is set, the non-diagonal terms in \eqref{eq.main} are neglected as in item 2 of Section \ref{sec.approx}.
  \item
  If the penultimate-significant bit is set, the terms in \eqref{eq.main} of order $\Omega^2$ are dropped as in item 1 of Section \ref{sec.approx}, which
   means the two $\beta$-coefficients \eqref{eq.betax2} and \eqref{eq.betay2} are replaced by zeros.
  \end{enumerate}
  These approximations may be accumulated by using an integer which has more than one bit set.
\end{itemize}
The output is a list of comments (starting lines with \texttt{\#}), and
snapshots of the trajectory, one per line. A snapshot line contains white-space separated
values from left to right:
\begin{enumerate}
\item
time $t$ since start of the trajectory, in seconds
\item
the coordinate triple of the position $x$, $y$, $z$ in the local horizontal system in meters
\item
the coordinate triple of the position $X$, $Y$, $Z$ in the inertial frame in meters
\item
the excursion angle $\varphi$ in the local horizontal system in degrees
\item
the azimuth angle $\lambda$ in the local horizontal system in degrees
\item
the value $\sqrt{\dot x^2+\dot y^2+\dot z^2}$ of the velocity measured in the local horizontal system in meters per second.
\end{enumerate}
There are lines that start with \texttt{\# P } dispersed in the output
that indicate transits when $\dot x=0$, obtained by quadratic interpolation
of three points on the trajectory 
where $\dot x$ changes sign. After \texttt{\# P} these
lines contain an integer count of this transit, the time when $\dot x$ was
zero in seconds, and an estimate of the frequency of the pendulum 
in Hz taking into account that two such transits occur per period.

\bibliography{all}

\begin{thebibliography}{30}%
\makeatletter
\providecommand \@ifxundefined [1]{%
 \@ifx{#1\undefined}
}%
\providecommand \@ifnum [1]{%
 \ifnum #1\expandafter \@firstoftwo
 \else \expandafter \@secondoftwo
 \fi
}%
\providecommand \@ifx [1]{%
 \ifx #1\expandafter \@firstoftwo
 \else \expandafter \@secondoftwo
 \fi
}%
\providecommand \natexlab [1]{#1}%
\providecommand \enquote  [1]{``#1''}%
\providecommand \bibnamefont  [1]{#1}%
\providecommand \bibfnamefont [1]{#1}%
\providecommand \citenamefont [1]{#1}%
\providecommand \href@noop [0]{\@secondoftwo}%
\providecommand \href [0]{\begingroup \@sanitize@url \@href}%
\providecommand \@href[1]{\@@startlink{#1}\@@href}%
\providecommand \@@href[1]{\endgroup#1\@@endlink}%
\providecommand \@sanitize@url [0]{\catcode `\\12\catcode `\$12\catcode
  `\&12\catcode `\#12\catcode `\^12\catcode `\_12\catcode `\%12\relax}%
\providecommand \@@startlink[1]{}%
\providecommand \@@endlink[0]{}%
\providecommand \url  [0]{\begingroup\@sanitize@url \@url }%
\providecommand \@url [1]{\endgroup\@href {#1}{\urlprefix }}%
\providecommand \urlprefix  [0]{URL }%
\providecommand \Eprint [0]{\href }%
\providecommand \doibase [0]{https://doi.org/}%
\providecommand \selectlanguage [0]{\@gobble}%
\providecommand \bibinfo  [0]{\@secondoftwo}%
\providecommand \bibfield  [0]{\@secondoftwo}%
\providecommand \translation [1]{[#1]}%
\providecommand \BibitemOpen [0]{}%
\providecommand \bibitemStop [0]{}%
\providecommand \bibitemNoStop [0]{.\EOS\space}%
\providecommand \EOS [0]{\spacefactor3000\relax}%
\providecommand \BibitemShut  [1]{\csname bibitem#1\endcsname}%
\let\auto@bib@innerbib\@empty
\bibitem [{\citenamefont {Noble}(1952)}]{NobleAJP20}%
  \BibitemOpen
  \bibfield  {author} {\bibinfo {author} {\bibfnamefont {W.~J.}\ \bibnamefont
  {Noble}},\ }\bibfield  {title} {\bibinfo {title} {A direct treatment of the
  foucault pendulum},\ }\href {https://doi.org/10.1119/1.1933230} {\bibfield
  {journal} {\bibinfo  {journal} {Am. J. Phys.}\ }\textbf {\bibinfo {volume}
  {20}},\ \bibinfo {pages} {334} (\bibinfo {year} {1952})}\BibitemShut
  {NoStop}%
\bibitem [{\citenamefont {Rapp}(1991)}]{RappGG1}%
  \BibitemOpen
  \bibfield  {author} {\bibinfo {author} {\bibfnamefont {R.~H.}\ \bibnamefont
  {Rapp}},\ }\href@noop {} {\emph {\bibinfo {title} {Geometric Geodesy Part
  1}}}\ (\bibinfo  {publisher} {Ohio State University},\ \bibinfo {address}
  {Columbus, Ohio},\ \bibinfo {year} {1991})\BibitemShut {NoStop}%
\bibitem [{\citenamefont {{National Imagery and Mapping
  Agency}}(2000)}]{NIMA8350}%
  \BibitemOpen
  \bibfield  {author} {\bibinfo {author} {\bibnamefont {{National Imagery and
  Mapping Agency}}},\ }\href
  {https://gis-lab.info/docs/nima-tr8350.2-wgs84fin.pdf} {\emph {\bibinfo
  {title} {Department {O}f {D}efense {W}orld {G}eodetic {S}ystem 1984}}},\
  \bibinfo {type} {Tech. Rep.}\ \bibinfo {number} {TR8350.2}\ (\bibinfo
  {institution} {{NIMA}},\ \bibinfo {year} {2000})\BibitemShut {NoStop}%
\bibitem [{\citenamefont {Vermeille}(2002)}]{VermeilleJG76}%
  \BibitemOpen
  \bibfield  {author} {\bibinfo {author} {\bibfnamefont {H.}~\bibnamefont
  {Vermeille}},\ }\bibfield  {title} {\bibinfo {title} {Direct transformation
  from geocentric coordinates to geodetic coordinates},\ }\href
  {https://doi.org/10.1007/s00190-002-0273-6} {\bibfield  {journal} {\bibinfo
  {journal} {J. Geod.}\ }\textbf {\bibinfo {volume} {76}},\ \bibinfo {pages}
  {451} (\bibinfo {year} {2002})}\BibitemShut {NoStop}%
\bibitem [{\citenamefont {Mathar}(2019)}]{MatharVixra1909}%
  \BibitemOpen
  \bibfield  {author} {\bibinfo {author} {\bibfnamefont {R.~J.}\ \bibnamefont
  {Mathar}},\ }\bibfield  {title} {\bibinfo {title} {The non-forced spherical
  pendulum: semi-numerical solutions},\ }\bibfield  {journal} {\bibinfo
  {journal} {vixra:1909.0201}\ }\href {https://doi.org/10.5281/zenodo.7636422}
  {10.5281/zenodo.7636422} (\bibinfo {year} {2019})\BibitemShut {NoStop}%
\bibitem [{\citenamefont {de~Icaza-Herrera}\ and\ \citenamefont
  {Casta{\~n}o}(2011)}]{IcazaAM218}%
  \BibitemOpen
  \bibfield  {author} {\bibinfo {author} {\bibfnamefont {M.}~\bibnamefont
  {de~Icaza-Herrera}}\ and\ \bibinfo {author} {\bibfnamefont {V.~M.}\
  \bibnamefont {Casta{\~n}o}},\ }\bibfield  {title} {\bibinfo {title}
  {Generalized lagrangian of the parametric foucault pendulum with dissipative
  forces},\ }\href {https://doi.org/10.1007/s00707-010-0392-8} {\bibfield
  {journal} {\bibinfo  {journal} {Acta Mech.}\ }\textbf {\bibinfo {volume}
  {218}},\ \bibinfo {pages} {45} (\bibinfo {year} {2011})}\BibitemShut
  {NoStop}%
\bibitem [{\citenamefont {Boulanger}\ and\ \citenamefont
  {Buisseret}(2020)}]{BoulangerPhys2}%
  \BibitemOpen
  \bibfield  {author} {\bibinfo {author} {\bibfnamefont {N.}~\bibnamefont
  {Boulanger}}\ and\ \bibinfo {author} {\bibfnamefont {F.}~\bibnamefont
  {Buisseret}},\ }\bibfield  {title} {\bibinfo {title} {The formulation of
  classical mechanics with foucault's pendulum},\ }\href
  {https://doi.org/10.3390/physics2040030} {\bibfield  {journal} {\bibinfo
  {journal} {Physics}\ }\textbf {\bibinfo {volume} {2}},\ \bibinfo {pages}
  {531} (\bibinfo {year} {2020})}\BibitemShut {NoStop}%
\bibitem [{\citenamefont {Jos\'e}\ and\ \citenamefont
  {Saletan}(1998)}]{JoseCM}%
  \BibitemOpen
  \bibinfo {editor} {\bibfnamefont {J.~V.}\ \bibnamefont {Jos\'e}}\ and\
  \bibinfo {editor} {\bibfnamefont {E.~J.}\ \bibnamefont {Saletan}},\ eds.,\
  \href@noop {} {\emph {\bibinfo {title} {Classical Mechanics: a contemporary
  approach}}}\ (\bibinfo  {publisher} {Cambridge University Press},\ \bibinfo
  {address} {Cambridge, UK},\ \bibinfo {year} {1998})\BibitemShut {NoStop}%
\bibitem [{\citenamefont {Somerville}(1972)}]{SomervilleQJRAS13}%
  \BibitemOpen
  \bibfield  {author} {\bibinfo {author} {\bibfnamefont {W.~B.}\ \bibnamefont
  {Somerville}},\ }\bibfield  {title} {\bibinfo {title} {The description of
  foucault's pendulum},\ }\href
  {https://adsabs.harvard.edu/abs/1972QJRAS..13...40S} {\bibfield  {journal}
  {\bibinfo  {journal} {Quart. J. R. Astron. Soc.}\ }\textbf {\bibinfo {volume}
  {13}},\ \bibinfo {pages} {40} (\bibinfo {year} {1972})}\BibitemShut {NoStop}%
\bibitem [{\citenamefont {Cartmell}\ \emph {et~al.}(2020)\citenamefont
  {Cartmell}, \citenamefont {Faller}, \citenamefont {Lockerbie},\ and\
  \citenamefont {Handous}}]{CartmellPRSA2019}%
  \BibitemOpen
  \bibfield  {author} {\bibinfo {author} {\bibfnamefont {M.~P.}\ \bibnamefont
  {Cartmell}}, \bibinfo {author} {\bibfnamefont {J.~E.}\ \bibnamefont
  {Faller}}, \bibinfo {author} {\bibfnamefont {N.~A.}\ \bibnamefont
  {Lockerbie}},\ and\ \bibinfo {author} {\bibfnamefont {E.}~\bibnamefont
  {Handous}},\ }\bibfield  {title} {\bibinfo {title} {On the modelling and
  testing of a laboratory-scale foucault pendulum as a precursor for the design
  of a high-performance measurement instrument},\ }\href
  {https://doi.org/10.1098/rspa.2019.0680} {\bibfield  {journal} {\bibinfo
  {journal} {Proc. Royal Soc. A}\ }\textbf {\bibinfo {volume} {476}},\ \bibinfo
  {pages} {20190680} (\bibinfo {year} {2020})}\BibitemShut {NoStop}%
\bibitem [{\citenamefont {Cartmell}\ \emph {et~al.}(2021)\citenamefont
  {Cartmell}, \citenamefont {Lockerbie},\ and\ \citenamefont
  {Faller}}]{CartmellNCPS2021}%
  \BibitemOpen
  \bibfield  {author} {\bibinfo {author} {\bibfnamefont {M.~P.}\ \bibnamefont
  {Cartmell}}, \bibinfo {author} {\bibfnamefont {N.~A.}\ \bibnamefont
  {Lockerbie}},\ and\ \bibinfo {author} {\bibfnamefont {J.~E.}\ \bibnamefont
  {Faller}},\ }\bibfield  {title} {\bibinfo {title} {Towards a high-performance
  foucault pendulum},\ }in\ \href
  {https://doi.org/10.1007/978-3-030-81166-2_31} {\emph {\bibinfo {booktitle}
  {Advances in Nonlinear Dynamics}}},\ \bibinfo {series and number} {NODYCON
  Conference Proceedings},\ \bibinfo {editor} {edited by\ \bibinfo {editor}
  {\bibfnamefont {W.}~\bibnamefont {Lacarbonara}}, \bibinfo {editor}
  {\bibfnamefont {B.}~\bibnamefont {Balachandran}}, \bibinfo {editor}
  {\bibfnamefont {M.~J.}\ \bibnamefont {Leamy}}, \bibinfo {editor}
  {\bibfnamefont {J.}~\bibnamefont {Ma}}, \emph {et~al.}}\ (\bibinfo
  {publisher} {Springer},\ \bibinfo {year} {2021})\ pp.\ \bibinfo {pages}
  {343--353}\BibitemShut {NoStop}%
\bibitem [{\citenamefont {Pippard}(1988)}]{PippardPRSL420}%
  \BibitemOpen
  \bibfield  {author} {\bibinfo {author} {\bibfnamefont {A.~B.}\ \bibnamefont
  {Pippard}},\ }\bibfield  {title} {\bibinfo {title} {The parametrically
  maintained foucault pendulum and its perturbations},\ }\href
  {https://doi.org/10.1098/rspa.1988.0118} {\bibfield  {journal} {\bibinfo
  {journal} {Proc. R. Soc. Lond. A}\ }\textbf {\bibinfo {volume} {420}},\
  \bibinfo {pages} {81} (\bibinfo {year} {1988})}\BibitemShut {NoStop}%
\bibitem [{\citenamefont {Persson}(2015)}]{PerssonQJRMS141}%
  \BibitemOpen
  \bibfield  {author} {\bibinfo {author} {\bibfnamefont {A.}~\bibnamefont
  {Persson}},\ }\bibfield  {title} {\bibinfo {title} {Is the coriolis effect an
  `optical illusion'?},\ }\href {https://doi.org/10.1002/qj.2477} {\bibfield
  {journal} {\bibinfo  {journal} {Quart. J. Roy. Meteor. Soc.}\ }\textbf
  {\bibinfo {volume} {141}},\ \bibinfo {pages} {1957} (\bibinfo {year}
  {2015})}\BibitemShut {NoStop}%
\bibitem [{\citenamefont {Phillips}(2000)}]{PhillipsBAMS81}%
  \BibitemOpen
  \bibfield  {author} {\bibinfo {author} {\bibfnamefont {N.~A.}\ \bibnamefont
  {Phillips}},\ }\bibfield  {title} {\bibinfo {title} {An explication of the
  coriolis effect},\ }\href
  {https://doi.org/10.1175/1520-0477(2000)081<0299:AEOTCE>2.3.CO;2} {\bibfield
  {journal} {\bibinfo  {journal} {Bull. Am. Meteor. Soc.}\ }\textbf {\bibinfo
  {volume} {81}},\ \bibinfo {pages} {299} (\bibinfo {year} {2000})}\BibitemShut
  {NoStop}%
\bibitem [{\citenamefont {Durran}(1993)}]{DurranAMS74}%
  \BibitemOpen
  \bibfield  {author} {\bibinfo {author} {\bibfnamefont {D.~R.}\ \bibnamefont
  {Durran}},\ }\bibfield  {title} {\bibinfo {title} {Is the coriolis force
  really responsible for the inertial oscillation?},\ }\href
  {https://doi.org/10.1175/1520-0477(1993)074<2179:ITCFRR>2.0.CO;2} {\bibfield
  {journal} {\bibinfo  {journal} {Bull. Am. Metor. Soc.}\ }\textbf {\bibinfo
  {volume} {74}},\ \bibinfo {pages} {2179} (\bibinfo {year}
  {1993})}\BibitemShut {NoStop}%
\bibitem [{\citenamefont {Desloge}\ and\ \citenamefont
  {Karch}(1977)}]{DeslogeAJP45}%
  \BibitemOpen
  \bibfield  {author} {\bibinfo {author} {\bibfnamefont {E.~A.}\ \bibnamefont
  {Desloge}}\ and\ \bibinfo {author} {\bibfnamefont {R.~I.}\ \bibnamefont
  {Karch}},\ }\bibfield  {title} {\bibinfo {title} {Noether's theorem in
  classical mechanics},\ }\href {https://doi.org/10.1119/1.10616} {\bibfield
  {journal} {\bibinfo  {journal} {Am. J. Phys.}\ }\textbf {\bibinfo {volume}
  {45}},\ \bibinfo {pages} {336} (\bibinfo {year} {1977})}\BibitemShut
  {NoStop}%
\bibitem [{\citenamefont {Krivoruchenko}(2009)}]{KrivoruPU52}%
  \BibitemOpen
  \bibfield  {author} {\bibinfo {author} {\bibfnamefont {M.~I.}\ \bibnamefont
  {Krivoruchenko}},\ }\bibfield  {title} {\bibinfo {title} {Rotation of the
  swing plane of foucault's pendulum and thomas spin precession: two sides of
  one coin},\ }\href {https://doi.org/10.3367/UFNe.0179.200908e.0873}
  {\bibfield  {journal} {\bibinfo  {journal} {Physics-Uspekhi}\ }\textbf
  {\bibinfo {volume} {52}},\ \bibinfo {pages} {821} (\bibinfo {year}
  {2009})}\BibitemShut {NoStop}%
\bibitem [{\citenamefont {MacMillan}(1915)}]{MacMillanAJM37}%
  \BibitemOpen
  \bibfield  {author} {\bibinfo {author} {\bibfnamefont {W.~D.}\ \bibnamefont
  {MacMillan}},\ }\bibfield  {title} {\bibinfo {title} {On foucault's
  pendulum},\ }\href {https://doi.org/10.2307/2370259} {\bibfield  {journal}
  {\bibinfo  {journal} {Am. J. Math.}\ }\textbf {\bibinfo {volume} {37}},\
  \bibinfo {pages} {95} (\bibinfo {year} {1915})}\BibitemShut {NoStop}%
\bibitem [{\citenamefont {Schulz-DuBois}(1970)}]{SchulzAJP38}%
  \BibitemOpen
  \bibfield  {author} {\bibinfo {author} {\bibfnamefont {E.~O.}\ \bibnamefont
  {Schulz-DuBois}},\ }\bibfield  {title} {\bibinfo {title} {Foucault pendulum
  experiment by kamerlingh onnes and degenerate perturbation theory},\ }\href
  {https://doi.org/10.1119/1.1976270} {\bibfield  {journal} {\bibinfo
  {journal} {Am. J. Phys.}\ }\textbf {\bibinfo {volume} {38}},\ \bibinfo
  {pages} {173} (\bibinfo {year} {1970})}\BibitemShut {NoStop}%
\bibitem [{\citenamefont {Das}\ \emph {et~al.}(2002)\citenamefont {Das},
  \citenamefont {Talukdar},\ and\ \citenamefont {Shamanna}}]{DasCJP52}%
  \BibitemOpen
  \bibfield  {author} {\bibinfo {author} {\bibfnamefont {U.}~\bibnamefont
  {Das}}, \bibinfo {author} {\bibfnamefont {B.}~\bibnamefont {Talukdar}},\ and\
  \bibinfo {author} {\bibfnamefont {J.}~\bibnamefont {Shamanna}},\ }\bibfield
  {title} {\bibinfo {title} {Indirect analytic representations of foucault's
  pendulum},\ }\href {https://doi.org/10.1023/A:1021819627736} {\bibfield
  {journal} {\bibinfo  {journal} {Czech. J. Phys.}\ }\textbf {\bibinfo {volume}
  {52}},\ \bibinfo {pages} {1321} (\bibinfo {year} {2002})}\BibitemShut
  {NoStop}%
\bibitem [{\citenamefont {Condurache}\ and\ \citenamefont
  {Martinusi}(2008)}]{ConduracheIJNLM43}%
  \BibitemOpen
  \bibfield  {author} {\bibinfo {author} {\bibfnamefont {D.}~\bibnamefont
  {Condurache}}\ and\ \bibinfo {author} {\bibfnamefont {V.}~\bibnamefont
  {Martinusi}},\ }\bibfield  {title} {\bibinfo {title} {Foucaul pendulum-like
  problems: A tensorial approach},\ }\href
  {https://doi.org/10.1016/j.ijnonlinmec.2008.03.009} {\bibfield  {journal}
  {\bibinfo  {journal} {Int. J. Non-Linear Mech.}\ }\textbf {\bibinfo {volume}
  {43}},\ \bibinfo {pages} {743} (\bibinfo {year} {2008})}\BibitemShut
  {NoStop}%
\bibitem [{\citenamefont {Bromwich}(1914)}]{BromwichPLMS13}%
  \BibitemOpen
  \bibfield  {author} {\bibinfo {author} {\bibfnamefont {T.~J.}\ \bibnamefont
  {Bromwich}},\ }\bibfield  {title} {\bibinfo {title} {The theory of foucault's
  pendulum},\ }\href {https://doi.org/10.1112/plms/s-13.1.222} {\bibfield
  {journal} {\bibinfo  {journal} {Proc. Lond. Math Soc.}\ }\textbf {\bibinfo
  {volume} {s2-13}},\ \bibinfo {pages} {222} (\bibinfo {year}
  {1914})}\BibitemShut {NoStop}%
\bibitem [{\citenamefont {Opat}(1991)}]{OpatAJP59}%
  \BibitemOpen
  \bibfield  {author} {\bibinfo {author} {\bibfnamefont {G.~I.}\ \bibnamefont
  {Opat}},\ }\bibfield  {title} {\bibinfo {title} {The precession fo a foucault
  pendulum viewed as a beat phenomenon of a conical pendulum subject to a
  coriolis force},\ }\href {https://doi.org/10.1119/1.16729} {\bibfield
  {journal} {\bibinfo  {journal} {Am. J. Phys.}\ }\textbf {\bibinfo {volume}
  {59}},\ \bibinfo {pages} {822} (\bibinfo {year} {1991})}\BibitemShut
  {NoStop}%
\bibitem [{\citenamefont {Sommeria}(2017)}]{SommeriaCRP18}%
  \BibitemOpen
  \bibfield  {author} {\bibinfo {author} {\bibfnamefont {J.}~\bibnamefont
  {Sommeria}},\ }\bibfield  {title} {\bibinfo {title} {Foucault and the
  rotation of the earth},\ }\href {https://doi.org/10.1016/j.crhy.2017.11.003}
  {\bibfield  {journal} {\bibinfo  {journal} {Comptes Rend. Phys.}\ }\textbf
  {\bibinfo {volume} {18}},\ \bibinfo {pages} {520} (\bibinfo {year}
  {2017})}\BibitemShut {NoStop}%
\bibitem [{\citenamefont {Chessin}(1895)}]{ChessinAJM17}%
  \BibitemOpen
  \bibfield  {author} {\bibinfo {author} {\bibfnamefont {A.~S.}\ \bibnamefont
  {Chessin}},\ }\bibfield  {title} {\bibinfo {title} {On foucault's pendulum},\
  }\href {https://doi.org/10.2307/2369710} {\bibfield  {journal} {\bibinfo
  {journal} {Am. J. Math.}\ }\textbf {\bibinfo {volume} {17}},\ \bibinfo
  {pages} {81} (\bibinfo {year} {1895})}\BibitemShut {NoStop}%
\bibitem [{\citenamefont {Engeln-M\"ullges}\ and\ \citenamefont
  {Reuter}(1981)}]{EngelnMullges}%
  \BibitemOpen
  \bibfield  {author} {\bibinfo {author} {\bibfnamefont {G.}~\bibnamefont
  {Engeln-M\"ullges}}\ and\ \bibinfo {author} {\bibfnamefont {F.}~\bibnamefont
  {Reuter}},\ }\href@noop {} {\emph {\bibinfo {title} {Formelsammlung zur
  numerischen Mathematik}}},\ \bibinfo {series} {BI Hochschultaschenb\"ucher}\
  No.\ \bibinfo {number} {106}\ (\bibinfo  {publisher} {Bibliograph. Inst.},\
  \bibinfo {year} {1981})\BibitemShut {NoStop}%
\bibitem [{\citenamefont {Rutishauser}(1960)}]{RutishauserNM2}%
  \BibitemOpen
  \bibfield  {author} {\bibinfo {author} {\bibfnamefont {H.}~\bibnamefont
  {Rutishauser}},\ }\bibfield  {title} {\bibinfo {title} {Bemerkung zur
  numerischen integration gew\"ohlicher differentialgleichungen n-ter
  ordnung},\ }\href {https://doi.org/10.1007/BF01386228} {\bibfield  {journal}
  {\bibinfo  {journal} {Num. Math.}\ }\textbf {\bibinfo {volume} {2}},\
  \bibinfo {pages} {263} (\bibinfo {year} {1960})}\BibitemShut {NoStop}%
\bibitem [{\citenamefont {Fehlberg}(1987)}]{FehlbergZAMM67}%
  \BibitemOpen
  \bibfield  {author} {\bibinfo {author} {\bibfnamefont {E.}~\bibnamefont
  {Fehlberg}},\ }\bibfield  {title} {\bibinfo {title} {Neue
  runge-kutta-nystr\"om formelpaare 3(4)-ter und 4(5)-ter ordnung f\"ur $\ddot
  x=f(t,x,\dot x)$},\ }\href {https://doi.org/10.1002/zamm.19870670803}
  {\bibfield  {journal} {\bibinfo  {journal} {Z. Angew. Math. Mech.}\ }\textbf
  {\bibinfo {volume} {67}},\ \bibinfo {pages} {367} (\bibinfo {year}
  {1987})}\BibitemShut {NoStop}%
\bibitem [{\citenamefont {Fehlberg}(1974)}]{FehlbergTRR432}%
  \BibitemOpen
  \bibfield  {author} {\bibinfo {author} {\bibfnamefont {E.}~\bibnamefont
  {Fehlberg}},\ }\href {https://ntrs.nasa.gov/citations/19740026877} {\emph
  {\bibinfo {title} {Classical seventh-, sixths-, and fifth-order
  Runge-Kutta-Nystr\"om Formulas with stepsize control for general second-order
  differential equations}}},\ \bibinfo {type} {Tech. Rep.}\ \bibinfo {number}
  {TR R-432}\ (\bibinfo  {institution} {NASA},\ \bibinfo {year}
  {1974})\BibitemShut {NoStop}%
\bibitem [{\citenamefont {Fehlberg}(1975)}]{FehlbergComp14}%
  \BibitemOpen
  \bibfield  {author} {\bibinfo {author} {\bibfnamefont {E.}~\bibnamefont
  {Fehlberg}},\ }\bibfield  {title} {\bibinfo {title} {Klassische
  runge-kutta-nystr\"om-formeln mit schrittweitenkontrolle f\"ur
  differentialgleichungen $\ddot x=f(t,x,\dot x)$},\ }\href
  {https://doi.org/10.1007/BF02253548} {\bibfield  {journal} {\bibinfo
  {journal} {Computing}\ }\textbf {\bibinfo {volume} {14}},\ \bibinfo {pages}
  {371} (\bibinfo {year} {1975})}\BibitemShut {NoStop}%
\end{thebibliography}%

\end{document}